\documentclass[conference]{IEEEtran}
\IEEEoverridecommandlockouts
\usepackage{cite}
\usepackage{amsmath,amssymb,amsfonts}
\usepackage{algorithmic}
\usepackage{graphicx}
\usepackage{textcomp}
\usepackage{xcolor}
\def\BibTeX{{\rm B\kern-.05em{\sc i\kern-.025em b}\kern-.08em
    T\kern-.1667em\lower.7ex\hbox{E}\kern-.125emX}}

\begin{document}


\title{AI Based Landscape Sensing Using Radio Signals }
\date{\today}
\author{\IEEEauthorblockN{Vijaya~Yajnanarayana\textsuperscript{1,*}, Dongdong~Huang\textsuperscript{2}, Deep~Shrestha\textsuperscript{1}, Yi Geng\textsuperscript{2}, Ali Behravan\textsuperscript{1}, Erik Dahlman\textsuperscript{1}}
  \textsuperscript{1}Ericsson Research\\
  \textsuperscript{2}Ericsson R\&D\\

  *Email: vijaya.yajnanarayana@ericsson.com}
\maketitle
\thispagestyle{empty}
\pagestyle{empty}

\maketitle

\begin{abstract}
In many sensing applications, typically radio signals are emitted by a radar and from the bounced reflections of the obstacles, inference about the environment is made. Even though radars  can be used to sense the landscapes around the user-equipment (UE) such as whether UE is in the forested region, inside buildings, etc., it  is not suitable in many wireless applications as many UEs does not have radars in them. Using radar will also increase the cost and power requirements on the UEs in applications requiring sensing of the landscapes. In this paper, we provide a mechanism where basestation (BS) is able to sense the UE's landscape  without the use of a radar. We propose an artificial intelligence (AI)  based approach with suitable choice of the features derived from the wireless channel to infer the landscape of the UEs. Results for the proposed methods when applied to  practical environments such as London city scenario yields a precision score of more than 95 percent.
\end{abstract}


\section{Introduction}
\label{sec:intro}

In order to support new and enhanced use cases such as  holographic communication, high precision manufacturing,  autonomous driving, etc., the next generation 6G wireless system need to satisfy stringent connectivity, throughput, latency and reliability requirement as suggested in \cite{strinati-2019}\cite{6G-flagship-book}. In order to support these,  the next generation wireless system will have wide bandwidth and potentially operate in high attenuation frequency bands. Natural consequence of this is a densely deployed network with high basestation (BS) density to ensure coverage and connectivity. These densely deployed BSs need to collaborate with each other to support 6G use cases.

In order to support the stringent KPIs of the 6G network (refer to Table 1 of \cite{strinati-2019}), not only the distributed nature of deployed nodes need to be exploited fully, but also it is imperative to find newer strategies for network adaptation to counter the UE's extremely dynamic environment in the next generation networks. One such strategy is to employ  sensing of UE environments in a distributed way and tailor the network adaptation actions accordingly \cite{bourdoux-2020-white-paper}. If a radar infrastructure is added on top of the communication infrastructure, then several aspects of the UE environment such as range, Doppler, etc. can be detected providing a rich representation of the UE environment. However, in a typical communication system, most of the UEs does not have a radar built into them. Therefore, in this paper, we propose a method to sense the macro environment of a UE using communication signals alone. Specifically, we propose methods to detect the landscape around the UE to answer the hypothesis questions such as  whether the UE on a street, is the UE surrounded by a forested area, etc.

Detecting the landscape around the UE can have significant benefits. In future 6G systems, it is very important for the BS to understand the UE's environment to tailor the signal for it to counter the extreme propagation condition described above. Knowing the landscape helps the UE to predict the likelihood of the handover and can aid in mobility management. It can also enhance the digital twin representation of the UE in a 6G network for improved inference and action by the network infrastructure to support the KPI requirements.

\subsection{Related Work}
There exist several works on environment sensing using dedicated sensors and radars in UEs. In these works, UEs (typically an autonomous vehicle), uses the synthetic aperture radar (SAR) to illuminate the environment and from the return signal an inference about the environment is made. For example, in \cite{wu-2009-autom-sar},  the authors use SAR to sense parking lots near the vicinity of the UE. Similarly, in \cite{gao-2021-mimo-sar}, authors use sensing infrastructure on the road together with the wireless communication for environment sensing. These approaches are not feasible for majority of the UEs in the network since they do not have the radars in them.  Another approach to the landscape sensing includes the estimation of the position using GPS sensor or using the dedicated positional reference signal to localize the UE and use the geographic information system (GIS) with high-resolution maps to draw inference on the landscape. However, these additional GIS data and high resolution maps may not be available at the BS.

\subsection{Contributions}

In all the above discussed landscape-sensing methods, UEs need external support from one or more entities such as  sensors, radars, dedicated signaling, GIS servers, etc. This limits  the applicability of these methods for all UEs under all scenarios. To the best of our knowledge, this is the first work which proposes a method where the BSs can infer about the UE's landscape without the need for any new sensor or radar infrastructure requirements. In this paper, we propose an AI-based method which employs a computationally efficient feature engineering and leverages the dense deployment of the future network to employ a distributed method to arrive at the inference on the landscape.

\begin{figure}[t]
  \centering
  \fbox{\includegraphics[width=2.5 in]{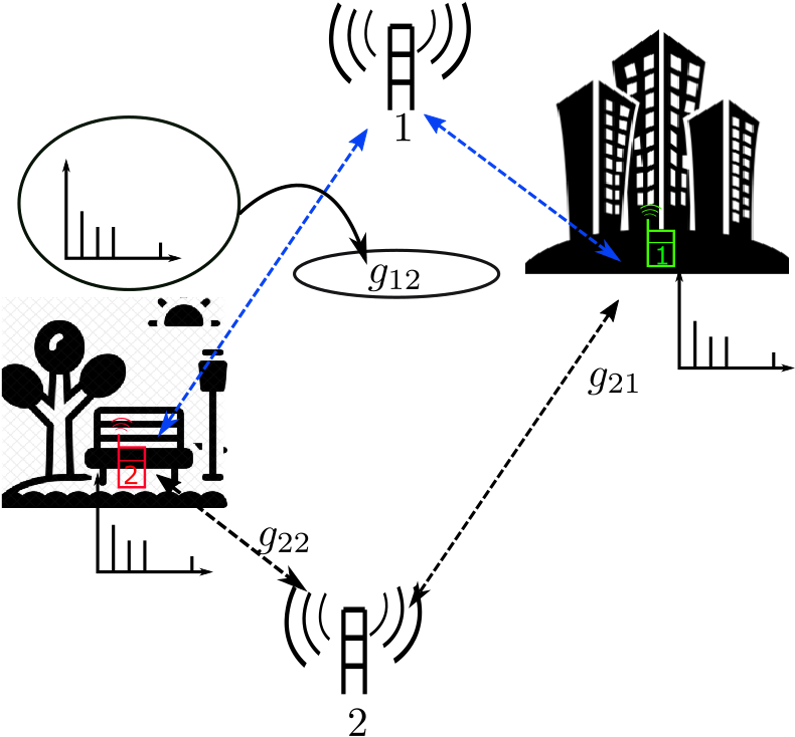}}

  \caption{Illustration of the UEs in forest and urban landscape, Two BSs  are serving them. By harvesting the path-gains from multiple BSs, the landscape around the UE can be detected.}
  \label{fig:dep1}
\end{figure}

\section{System design}
\label{sec:systemdesign}
We consider a typical deployment of the UEs  in a metro city like London, whose landscape could  include several geographical characteristics such as buildings, forests, streets, barren-landscapes, sea, etc. An illustration of such a deployment is as shown in the Fig~\ref{fig:dep1}, Here we show two UEs, UE-1 in an urban canyon and UE-2 in a forested area. The uplink (UL) impulse response having rich multi-path profiles carries information regarding the UE's  environment. Harvesting the UL impulse response from multiple base-stations can provide higher diversity due to the different geometric perspectives to the observations. For example, in Fig~\ref{fig:dep1}, the multi-path profiles from UE-1 to BS-1 and BS-2 will have a particular signature which will be different from multi-path signatures of UE-2 to BS-1 and BS-2. If the BSs co-ordinate by transferring  the multi-path profile of the UE to its serving BS\footnote{BS to BS communication is possible using 3GPP X2 interface}, then a better estimate about the landscape can be made.

There are several challenges with such a system design. Firstly, the  exchange of information between non-serving BSs to the serving BS for aggregating the  multi path-profiles of a UE  requires significant communication. Secondly, applying AI/ML methods on such large-dimension input features will be computationally expensive especially for an embedded BS hardware. To counter these challenges, we represent the multi-path profile of the link using a single large-scale channel statistic such as path-gain, a representation of this in the form of reference signal received power (RSRP)\footnote{RSRP and path-gains are interchangeably used}  can also be used. Since each BS will transfer only path-gain instead of entire multi-path profile, this reduces the input feature dimension for the estimation algorithm and also reduces the overhead in the distributed sensing setup. This is further improved by limiting the path-gain only to top $N$ strongest links. The path-gains between the links,  can also be obtained at the serving cell without any BS-BS communication. For example, by configuring the UE for periodic measurement reporting,  through radio resource control (RRC) signaling, the UE will report the neighbor-cell measurements at regular intervals which contains RSRPs (path-gains) to the neighbor cells and can be used for landscape detection.

Consider a typical deployment with $K$ BSs.  
A training set consisting of features and labels are first constructed by dropping the UEs at random positions at various landscapes and collecting the path-gains to various BSs and collating them into a set called training set. The $l$-th row of the such a training set is given by
\begin{equation}
  \label{eq:training-set1}
  T^{<l>} = \left\{ S_N\left(\mathcal{F}^{<l>}\right),\mathcal{L}^{<l>}\right\},
\end{equation}
where
\begin{equation}
  \label{eq:training-set2}
  \mathcal{F}^{<l>}=\left\{  g^{<l>}_{1}, g^{<l>}_{2}, \ldots, g^{<l>}_{K}\right\},
\end{equation}
\begin{equation}
  \label{eq:training-set3}
  \mathcal{L}^{<l>}=c
\end{equation}
with $g^{<l>}_{i}$ representing the path-gain to the $i$-th BS and $l\in \{1,\ldots,L\}$ denotes the $l$-th row of the training set and $c\in\mathcal{C}$, where ${\mathcal{C}}$ denotes the categories set consisting of landscape types such as street, forest, buildings, sea, etc. The function $S_N\left(\mbox{.}\right)$ retains the dominant $N$ paths while zeroing out the rest of the set, such that 
\begin{equation}
  \label{eq:card}
  \left| S_N\left(\mathcal{F}^{<l>} \right)\right| = N.
  \end{equation}
The $\left|\mbox{.}\right|$ denotes the cardinality of the set. 

\begin{figure}[t]
  \centering
   \includegraphics[width=2.5 in]{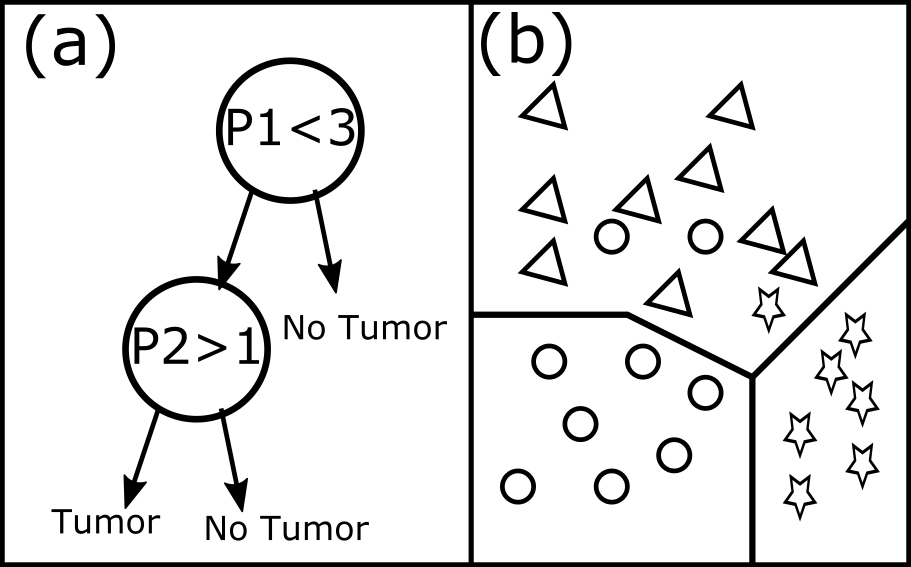}
  \caption{Decision tree and partition diagram of a random-forest classifier. (a) denotes a toy example of a decision tree with tests on two features on a graph to arrive at the decision on cancer. (b) The class boundaries in random-forest yields better generalization and less over-fitting due to the averaging of results from many decision trees.   }
  \label{fig:RandomForest}
\end{figure}

\section{AI Method}
The training set with appropriate features set design as discussed in the previous section is used to train a supervised  AI agent  for landscape estimation. Since the label belongs to a pre-defined set (called categories or landscape-types), The AI method should be a classification algorithm supervised or trained by this training set. Though several AI methods are possible, we believe that random-forest algorithm is well suited for this purpose, since the method is robust against outliers and perform well for highly non-linear partition boundaries. In the following, we discuss the random-forest algorithm briefly.

\subsection{Random-forest Algorithm}
Random-forest is an ensemble learner built on decision trees. A decision tree is a tree-like model with its nodes denoting a test on a feature and its branches denoting the consequence for a particular test outcome. Typically, the leaf-notes points to a class in a classification problem. An illustration of the decision tree  for a toy problem is shown in the Fig.~\ref{fig:RandomForest}(a).  Each step partitions the  training set and the process is repeated in a recursive manner until all the elements in the set has the same target class or there is no value in further partitioning. Practical classification problems have several features and typically need deep decision trees and they tend to over-fit the data resulting in poor performance on unseen data. To overcome this problem, ensemble classifier called random-forest  can be employed.  A random forest algorithm average over multiple decision trees and is shown to generalize the model better. An illustration of  the partition boundaries in a random-forest classifier is shown in the Fig~\ref{fig:RandomForest}(b) \cite{cutler-2012-random-forest}. In the next section, we will describe the network-deployment and simulation settings for the random-forest-based landscape detection algorithm.

\begin{figure}[t]
  \centering
    \includegraphics[width=1.9 in]{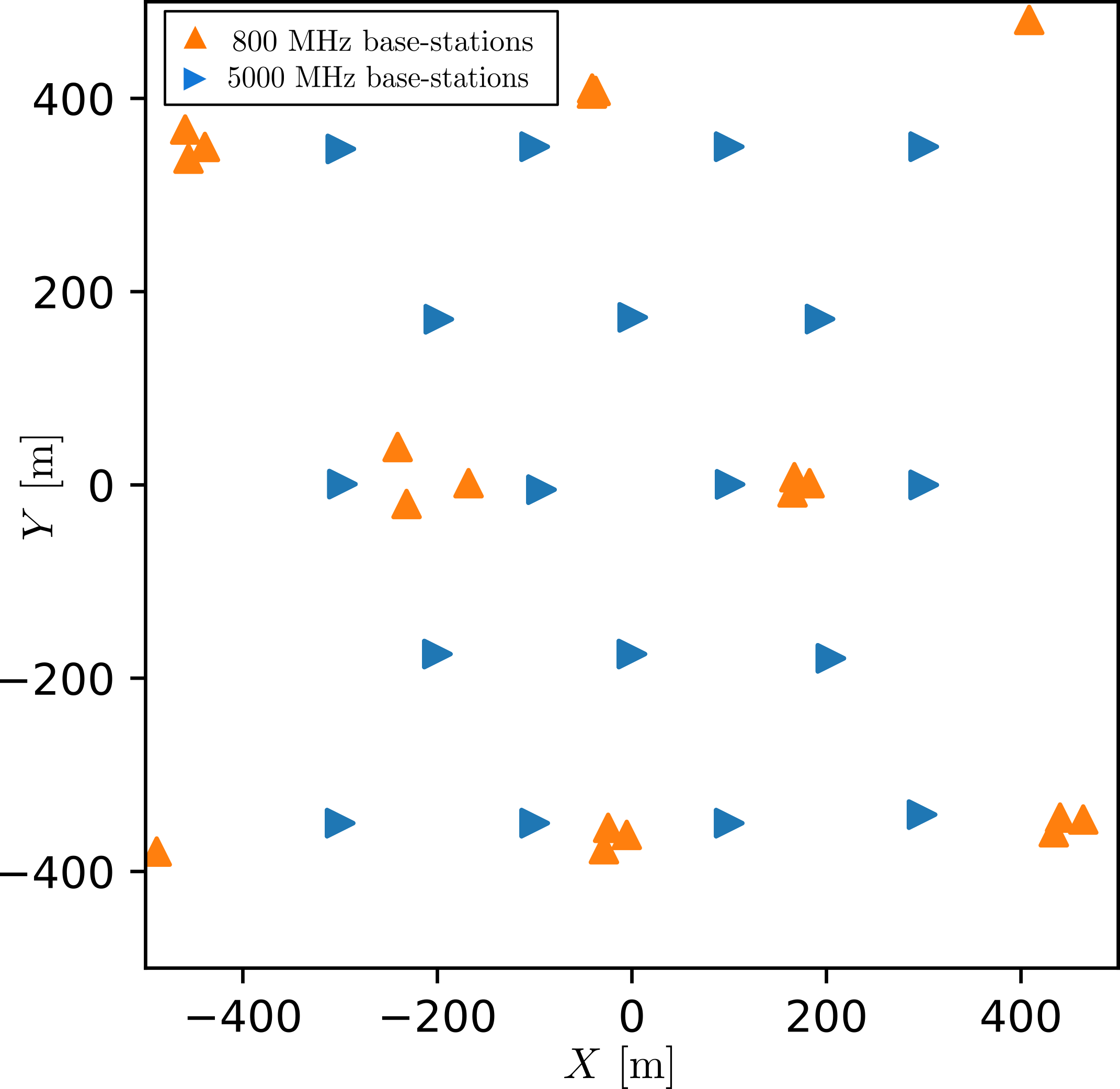}
  \caption{The multi-carrier deployment having $800$~MHz and $5$~GHz basestations covering $1\mbox{ km}^2$ section of central London.}
  \label{fig:cl}
\end{figure}

\begin{figure}[t]
  \centering
  \includegraphics[width=3.4 in]{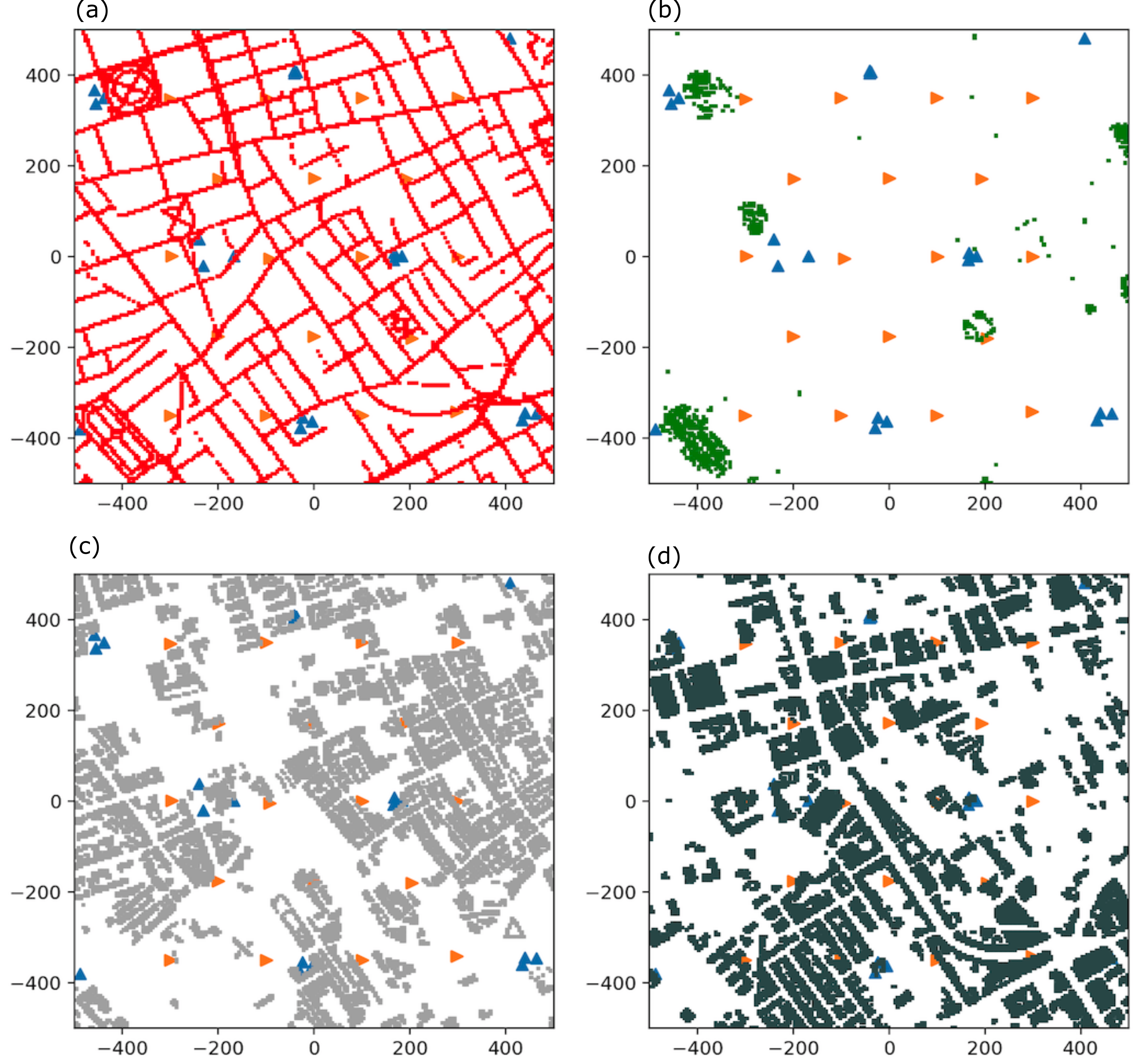}
  \caption{The four landscapes in the $\mbox{1 km x 1 km}$ section of central London. Sub figures (a), (b) (c) and (d) denotes the street, forest, building (8 -20 [m] height) and building (20-40 [m] height) respectively.  }
  \label{fig:cat}
\end{figure}

\section{ Network Deployment and simulation configuration} \label{sec:nd}
We use London city with its rich urban environment for accessing the landscape-detection performance of the proposed method. We consider a multi-carrier deployment of $ \mbox{1 km x 1 km}$ area of the central London. The region is served by two frequency layers $800$~MHz and $5$~GHz. These BSs are deployed as shown in the Fig~\ref{fig:cl}. The deployment has  $K=20$ BSs operating at $800$~MHz, and $K=54$ BSs ($18$ sites with $3$ sectors)  operating at $5$~GHz. At high frequency the signal attenuation is higher and hence require more BSs to provide connectivity.

There are several landscapes within the region of the Landon city that can be considered, Fig~\ref{fig:cat} shows forests, streets, buildings ($8\mbox{-}20\mbox{ [m]}$) and buildings ($20\mbox{-}40\mbox{ [m]}$). As explained in Section~\ref{sec:systemdesign}, the multi-path profiles and path gains experienced by the UE are influenced by the landscape the UE is in. We use a system simulation tool  which implements a simplified map based ray-tracing propagation for arriving at the path-gains at various UE drops on the region having the deployment as shown in Fig~\ref{fig:cl}.
\section{Results and Analysis} \label{sec:RA}
\subsection{Binary Hypothesis}\label{sec:BH}
Detection of whether the UE is in a particular type  of the landscape can be considered as a binary hypothesis problem. An example to these includes answers to  the questions such as ``whether the UE is on a street?'' , ``is the UE surrounded by a forested region?'', etc. Answers to these hypothesis questions can aid the network in taking appropriate actions to optimize performance. For example, detecting the UE on the street can aid in mobility actions such as handover, detecting the UE in forest can trigger action to extend coverage, etc.

We consider a system where the labels are binarized to $0$ and $1$ depending on whether the UE is in a particular type of landscape, i.e, in \eqref{eq:training-set3},
\begin{equation}
  \label{eq:bin}
  c = \begin{cases}
        1, & \mbox{If UE is in the landscape being detected}, \\
        0, &\mbox{Otherwise.}
        \end{cases}
 \end{equation}
      
 We consider two types of landscapes ``Streets'' and ``Forests''. A training set is constructed using a system simulation tool discussed above. A suit of training sets with distinct value of $N$, (refer to \eqref{eq:card})  each having $L=20000$ rows is constructed, and each set is used to separately  train the random-forest  algorithm  to assess its performance dependence on $N$. To assess the performance the trained AI agent is tested with a corresponding  validation set which is also comprised of $20000$ rows but is not seen by the AI agent before.
 \begin{figure*}[t]
   \centering
   \includegraphics[width = 4 in]{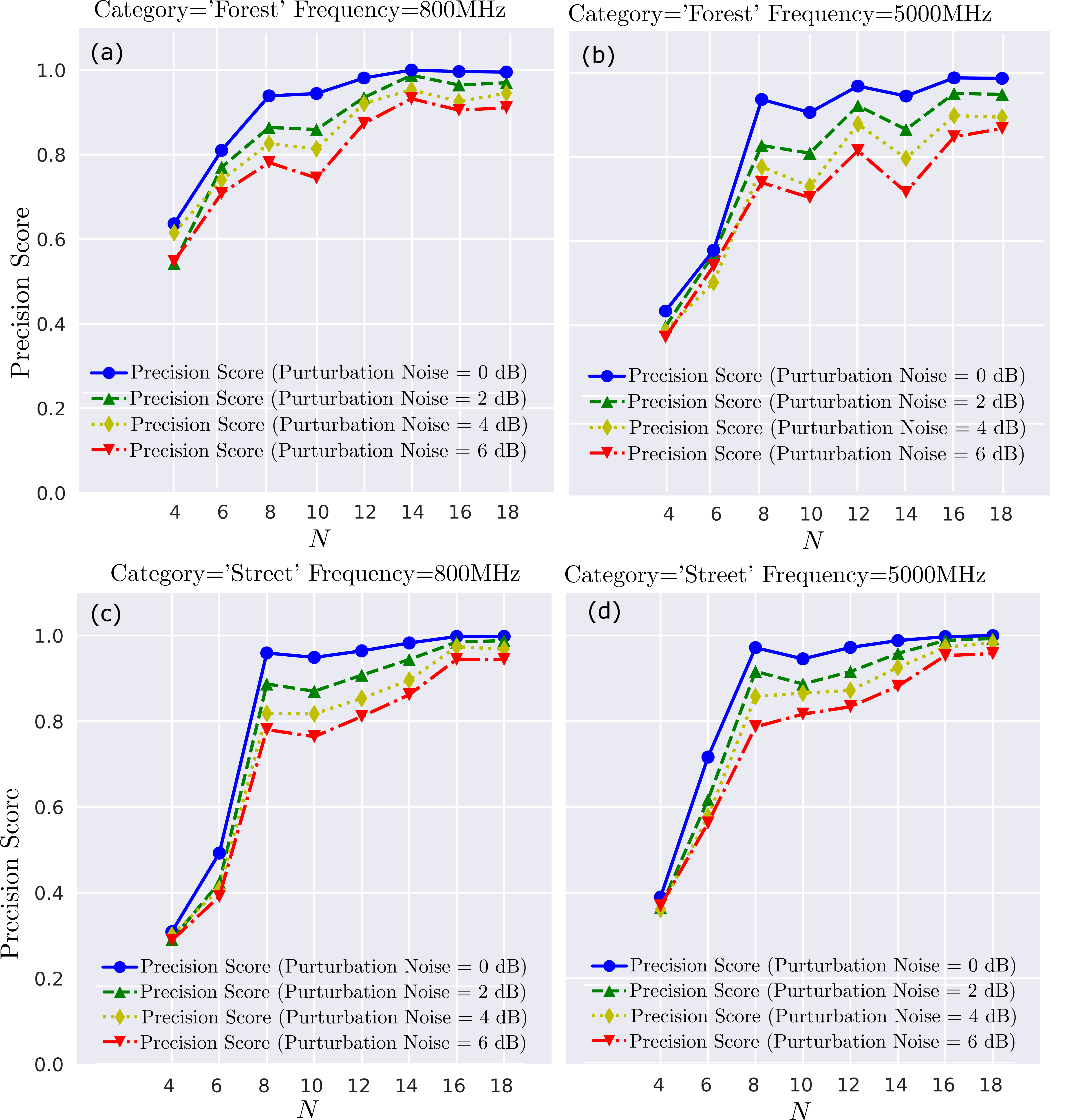}
   \caption{Performance of random-forest based AI method for 'Forest' and 'Street' landscape detection for a section of London city.}
   \label{fig:PrecisionBinaryHypothesis}
 \end{figure*}
 We define the performance in terms of precision and recall score. These are defined as
 \begin{equation}
   \label{eq:precision}
   \mbox{precision}=\frac{tp}{tp+fp}
 \end{equation}
 and
 \begin{equation}
   \label{eq:recall}
   \mbox{recall}=\frac{tp}{tp+fn}
 \end{equation}
 
 where $tp$, $fn$ and $fp$ are the true-positive, false-negative and false-positive values respectively. Precision score denotes the accuracy of the alternate-hypothesis (i.e., $c=1$) and recall score denotes the ratio of alternate hypothesis detected correctly by the agent, together they form the performance score for the detector. The precision score is more important than the recall score in the landscape sensing problem since  the hypothesis of UE in a particular landscape will trigger network  to take optimization action which can be catastrophic in case of a false alarm.

 The performance of the AI method for ``forest'' landscape detection is as shown in the Fig~\ref{fig:PrecisionBinaryHypothesis}(a) and Fig~\ref{fig:PrecisionBinaryHypothesis}(b) for $800$~MHz and $5$~GHz frequencies respectively. As expected the performance of the precision score improves with $N$, which is the number of strongest path-gains from diverse base-stations that is used by the AI agent in the hypothesis testing. 
 The perturbation noise is added to the validation set to  mimic the measurement uncertainty in the measurements. Typically, the agents are trained with the data from the measurement campaign. When the agent is deployed in the field for landscape detection, the measurements for the inference are derived from inexpensive RF transceivers in UEs or BSs (as explained in Section~\ref{sec:systemdesign}), which may not be highly accurate. The performance of the agent should be robust to such uncertainty, from the results shown in the Fig~\ref{fig:PrecisionBinaryHypothesis}, we notice that the estimators has sufficient margin against the measurement uncertainty. The similar performance analysis for the ``Street" landscape is shown in the Fig~\ref{fig:PrecisionBinaryHypothesis}(c,d).

 \begin{figure}[t]
   \centering
   \includegraphics[width =2.5 in]{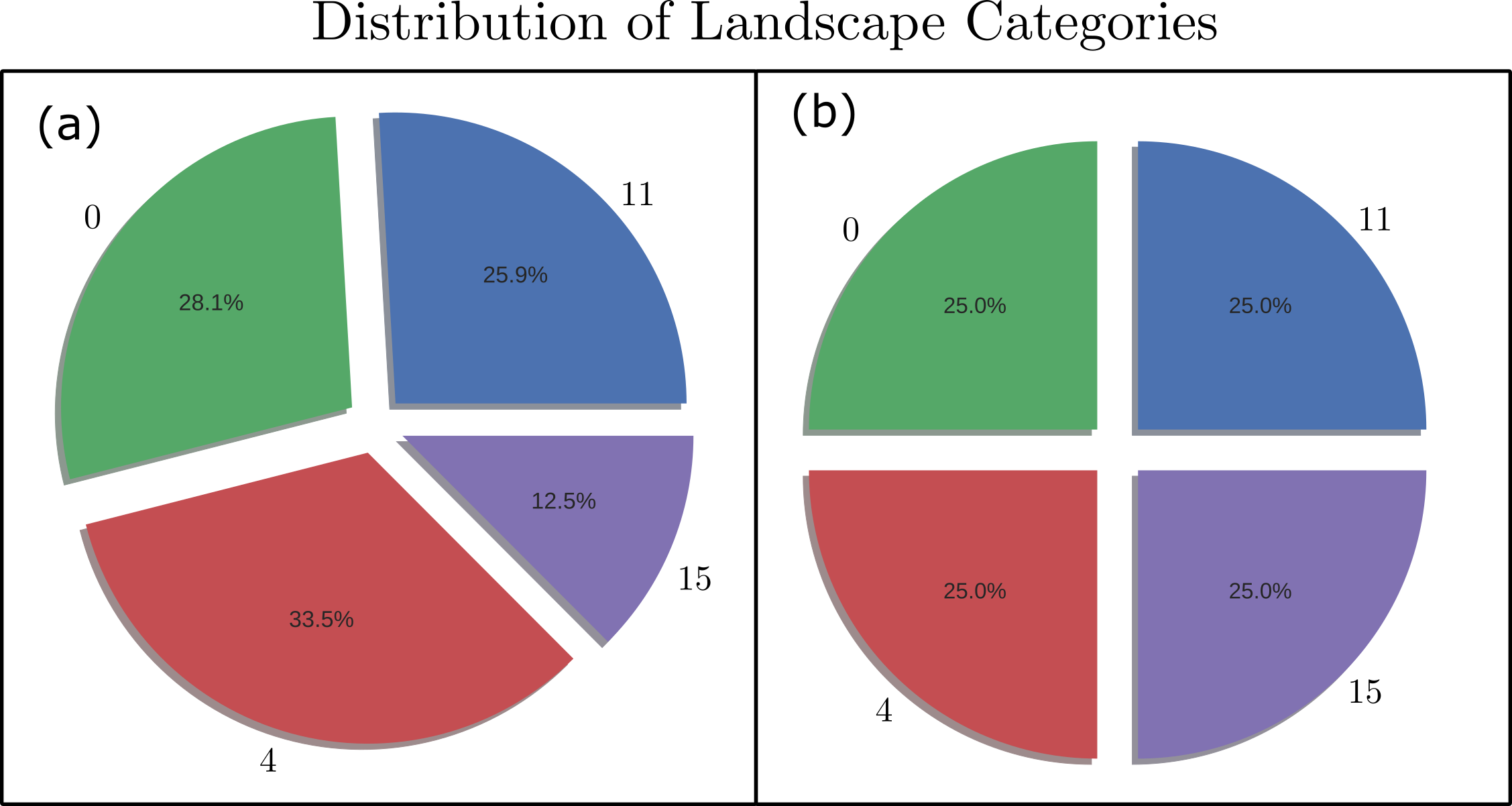}
   \caption{The ratio of landscape categories 11, 15, 4 and 0 for the section of London city considered for simulation. (a) Class-ratio as per the map. (b) Class-ratio after re-sampling the  features for balancing the classes.}
   \label{fig:classratio}
 \end{figure}
\subsection{Multiple Hypothesis}
\begin{figure}[t]
  \centering
  \fbox{\includegraphics[width= 3.4 in]{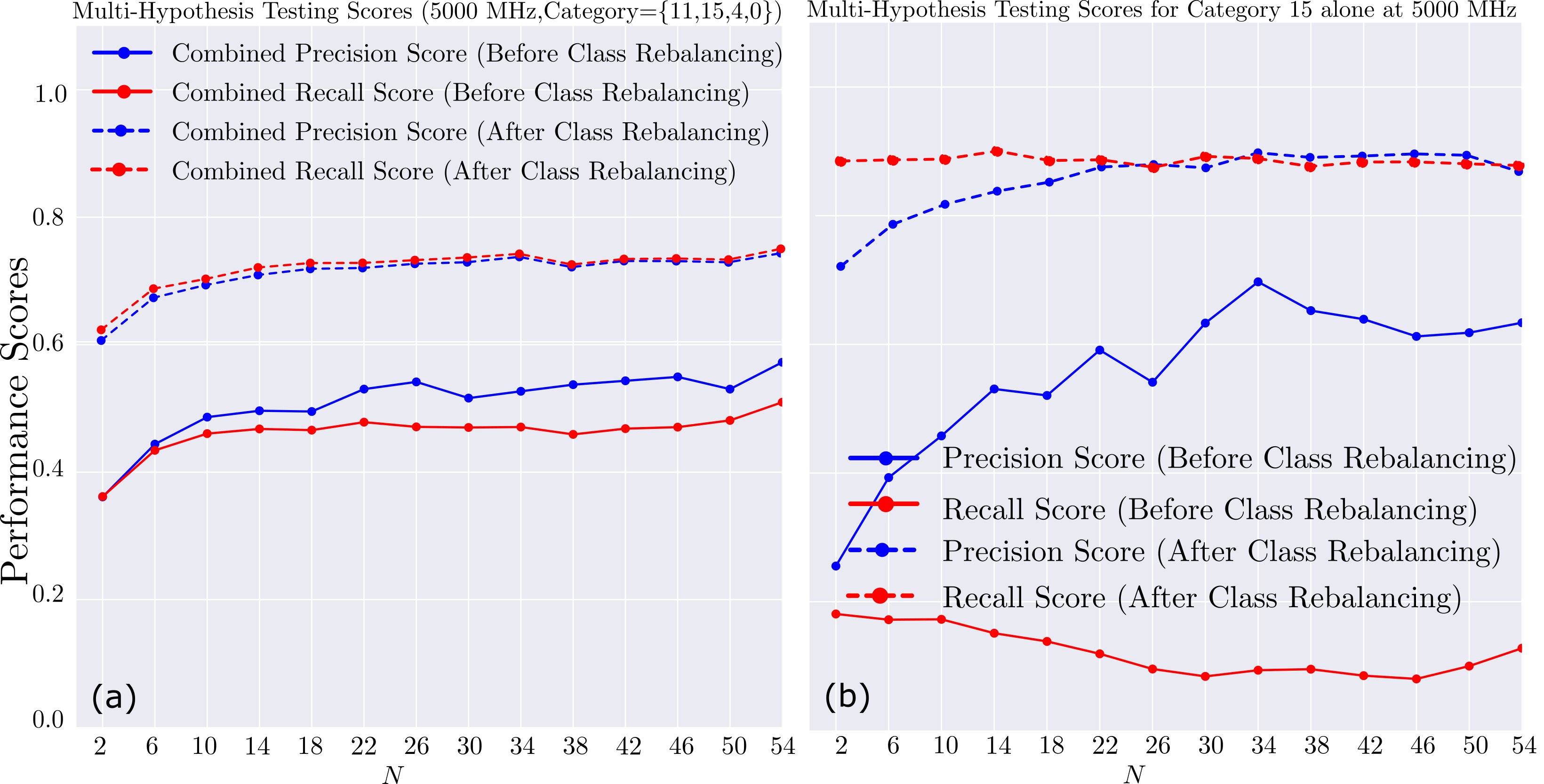}}
  \caption{Performance of the multi-hypothesis landscape sensing for $5$ GHz deployment (refer to Section~\ref{sec:nd}). Landscape category $11$, $15$, $4$ and $0$ is considered.  Class re-balancing by re-sampling the training set can improve performance scores. }
  \label{fig:MultiHypothesisScores}
\end{figure}

Below we discuss the multiple hypothesis scenario, where the objective is to identify  several landscape types together with the same feature set consisting of path-gain observations from multiple BSs. We consider landscape  categories $11$, $15$, and $4$ denoting UE in street, building and barren landscapes. Landscape category $0$  indicates that UE is in neither of the landscape categories $11$, $15$, and $4$. The proportion of these categories (class-ratios) play a crucial role in the performance of the detector. The Fig~\ref{fig:classratio}(a) shows the ratios of these regions in the section of the map of London city considered for the simulation. The averaged performance scores from all the categories for the multi-class detector categories is shown in the solid line plots of Fig~\ref{fig:MultiHypothesisScores}(a). The performance of the detector is improved by balancing the class ratio.  This can be done by re-sampling the training set in such a way that we have equal ratios for all the classes as shown in  Fig~\ref{fig:classratio}(b). The average performance scores of the landscape category detection after class-ratio re-balancing is shown in the broken-line plots of Fig~\ref{fig:MultiHypothesisScores}(a). The performance of the detector for the single class (street category alone) is shown in  Fig~\ref{fig:MultiHypothesisScores}(b). There is no noticeable improvement in the performance beyond ($N=10$), thus collaboration between 10 BSs in the dense deployment is sufficient to maximize detector performance. 

\begin{figure}[t]
  \centering
  \fbox{\includegraphics[width=3.4 in]{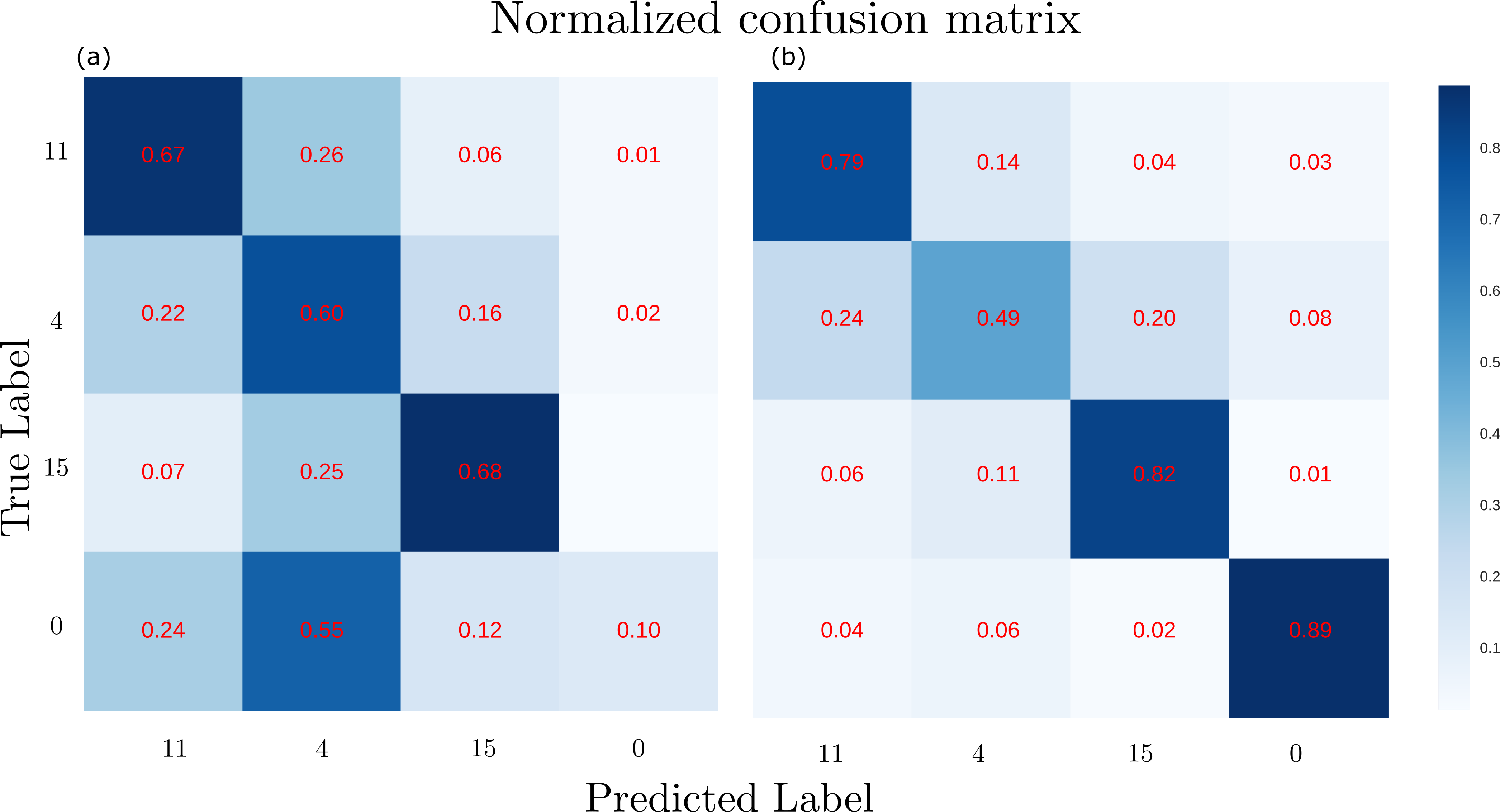}}
  \caption{The confusion matrices for $N=54$ and Frequency = $5$~GHz  AI model. (a) Before class-ratio re-balancing. (b) After class-ratio re-balancing.}
  \label{fig:ConfusionMatrices}
\end{figure}

The confusion matrices for before and after class-ratio re-balancing for $N=54$  case is as shown in the Fig~\ref{fig:ConfusionMatrices}(a) and Fig~\ref{fig:ConfusionMatrices}(b) respectively. The individual precision and recall scores computed from the confusion matrices by using \eqref{eq:precision} and \eqref{eq:recall} before and after class-ratio re-balancing  is shown in Table~\ref{tbl:multi-result}. Notice that after class-ratio re-balancing so that all categories are of equal ratio as shown in Fig~\ref{fig:classratio}(b), the average performance in terms of precision and recall scores has improved, however, the performance of some of the dominant landscape classes such as ``baron land'' has slightly deteriorated. Based on the application the class-ratios can be controlled in the training set to achieve the right trade-off in the multi-class detector's performance.

\begin{table}[t]
  \caption{Table for individual precision and recall scores for multi-hypothesis testing before and after class-ratio balancing.}
  \label{tbl:multi-result}
  \centering
  \begin{tabular}{ |p{1cm}|c|c|c|c|  }
 \hline
  &\multicolumn{2}{|c|}{Before class-ratio re-balancing} & \multicolumn{2}{|c|}{After class-ratio re-balancing } \\
 \hline
    Category  & Precision & Recall & Precision & Recall\\
 \hline
    Street & 0.68 & 0.66 & 0.82 & 0.76\\
    Baron & 0.60 & 0.36 & 0.49 &0.61 \\
    Building & 0.67 & 0.67 & 0.82 &0.7 \\
 \hline
\end{tabular}

\end{table}

\section{Conclusions}
Sensing the landscape around the UE can aid the network in taking appropriate actions to optimize performance. UEs fitted with radars can enable this, however, in future network only a small fraction of UEs will have this capability. An AI-based detector at the edge, which exploits the dense deployment of the future networks to arrive at the inference on the landscape sensing is promising. The feature engineering for such an AI detector to employ few strong path-gains from multiple  BSs will yield small-dimension input features and will reduce the computational complexity so that it can be implemented on an embedded BS hardware with minimum network overhead.

Results  in Fig~\ref{fig:PrecisionBinaryHypothesis}(a-d) shows that the proposed algorithm yields precision scores of greater than $95$\% for street and forest landscape. Perturbation analysis indicate that the proposed AI agent is robust to measurement uncertainty. Results from multi-hypothesis tests indicates that the performance of the detector depends on the class-ratios and can be improved by class-ratio re-balancing as shown in Fig~\ref{fig:MultiHypothesisScores}(a,b) . However, this process can slightly deteriorate performance of dominant landscape classes as shown in Table~\ref{tbl:multi-result}. Based on the application the class ratios can be controlled in the training set to achieve the right trade-off in the multi-class detector's performance.

\section*{Acknowledgements}
This work has been partly funded by the European Commission through the H2020 project Hexa-X (Grant Agreement no. 101015956)

\bibliography{main.bib}
\bibliographystyle{IEEE}

\end{document}